\documentclass[aps,prd,superscriptaddress,10pt,showpacs,notitlepage,  
nofootinbib
]{revtex4-1}

\usepackage{times}
\usepackage{color}
\usepackage{graphicx}

\begin{document}

\title{Pulsar glitches from quantum vortex networks}

\author{Giacomo Marmorini,$^{1}$  Shigehiro Yasui$^1$ and  Muneto Nitta,$^1$}
\date{$^1$Department of Physics $\&$ Research and Education Center for Natural Sciences, Keio University, Kanagawa 223-8521, Japan\\[2ex] 
}

\maketitle

{\bf
Neutron stars or pulsars are very rapidly rotating compact stars with extremely high density. 
One of the unsolved long-standing problems of these enigmatic celestial bodies is the origin of pulsars' glitches, {\it i.e.}, the sudden rapid deceleration in the rotation speed of neutron stars. Although many glitch events have been reported, there is no consensus on the microscopic mechanism responsible for them. One of the important characterizations of the glitches is the scaling law $P(E) \sim E^{-\alpha}$ of the probability distribution for a glitch with energy $E$. Here, we  reanalyse the accumulated up-to-date observation data to obtain the exponent $\alpha \approx 0.88$ for the scaling law, and propose a simple microscopic model that naturally deduces this scaling law without any free parameters. Our model explains the appearance of these glitches in terms of the presence 
of {\it quantum vortex networks} arising at the interface of two different kinds of superfluids in the core of neutron stars; a $p$-wave neutron superfluid in the inner core which interfaces with the $s$-wave neutron superfluid in the outer core, where each integer vortex in the $s$-wave superfluid connects to two half-quantized vortices in the $p$-wave superfluid through structures called ``boojums.'' 
}
\\

Neutron stars (NSs) and in particular pulsars 
are compact stars with the highest known density in our universe 
(about one solar mass within $10^3$ km$^3$) \cite{hewish68}, 
thereby providing an 
astrophysical laboratory to study phases of matter 
under extraordinary conditions:  
not only at very high density but also under rapid rotation and 
extremely strong magnetic fields  
(see refs.~\cite{Baym:2017whm,Graber:2016imq} for recent reviews). 
The study of NSs attracts great interest from researchers in diverse fields as recently there has been observations of highly massive NSs~\cite{Demorest:2010bx,Antoniadis1233232} and gravitational waves from a binary NS merger~\cite{TheLIGOScientific:2017qsa}.
One of the most important unsolved long-standing problems 
in NS physics is pulsar's glitches, {\it i.e.}, 
the abrupt deceleration of the rotation speed of NSs~\cite{reichley}. 
Although many glitch events have been reported, 
there is no consensus on the microscopic mechanism responsible for them. 
However, several ideas have been theoretically proposed: 
 starquakes~\cite{Baym1969,Pines1972}
 and the catastrophic unpinning of quantum superfluid vortices~\cite{Anderson:1975zze,Alpar1977}. 
 However, neither of these proposals has been rigorously proven or 
accepted widely 
in the community,  as explained below.

One of the important characterizations of the glitches is the scaling law of the cumulative probability distribution 
of glitch sizes 
$P(E) \sim E^{-\alpha}$ with an exponent $\alpha$, 
giving the probability of the occurrence of a glitch with energy $E$~\cite{morley-93} (see also ref.~\cite{Melatos_2009}).
The value of $\alpha \approx 0.14$ was obtained in ref.~\cite{morley-93}. 
Adopting the scaling law seems to be natural if one regards the glitches 
occurrence as consequences of starquakes in the crust region at the surface of NSs.
In fact, 
this scaling law resembles the Gutenberg-Richter law expressing the probability distribution of the total number of 
earthquakes of  certain magnitude~\cite{Gutenberg1}. 
This is known as an instance of power-law distributions
commonly found in diverse subjects in the natural, human and social sciences;
for example the Pareto distribution drawn from economics~\cite{Pareto}  
and scale-free networks such as the World Wide Web
in network science~\cite{Barabasi509}.
The ``network'' is one of the key ingredients in our study.

\begin{figure}[ht]
\begin{center}
\begin{tabular}{cc}
(a) Vortex (un)pinning & (b) Quantum vortex network\\
\includegraphics[scale=0.21]{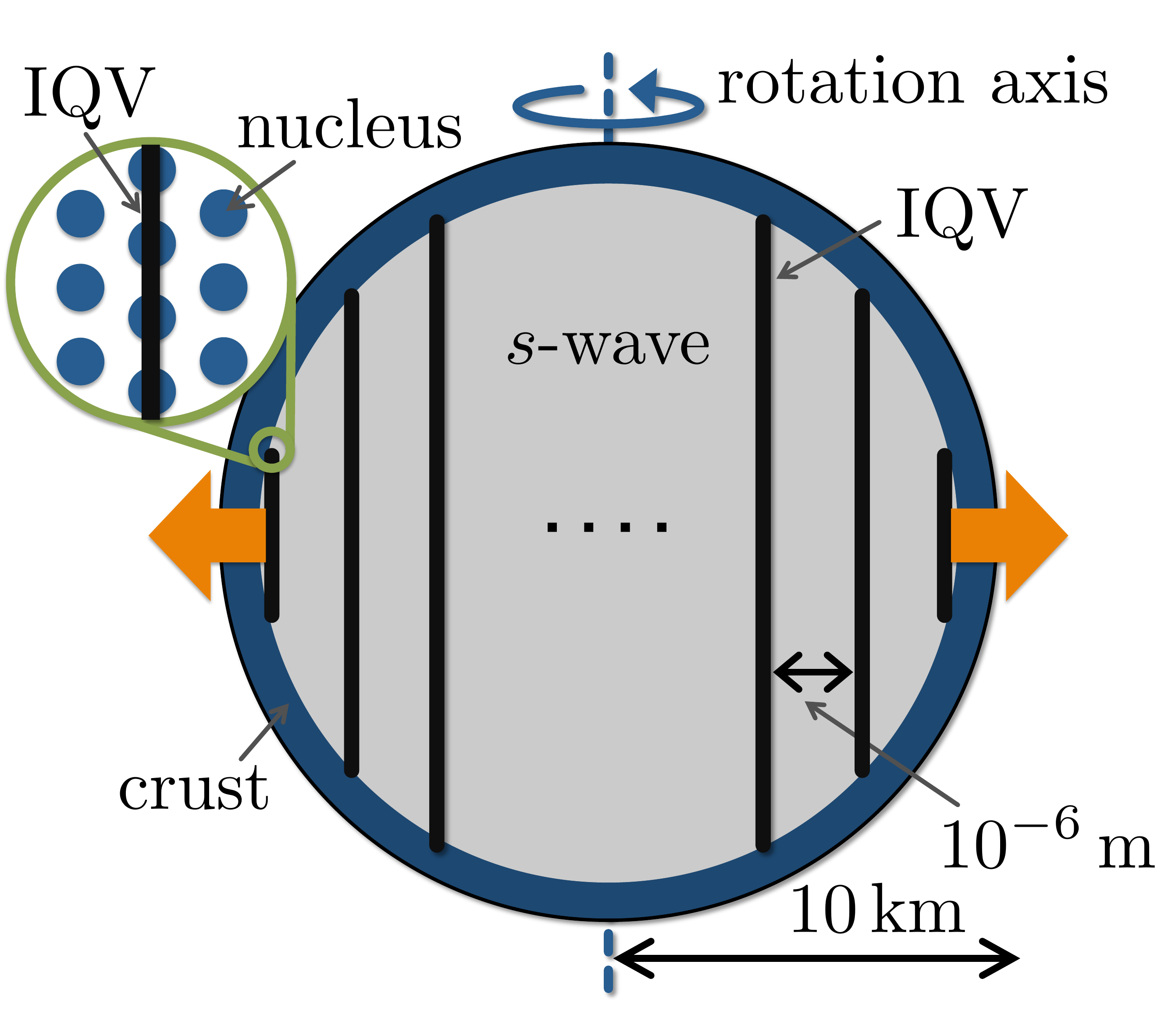} &
\includegraphics[scale=0.21]{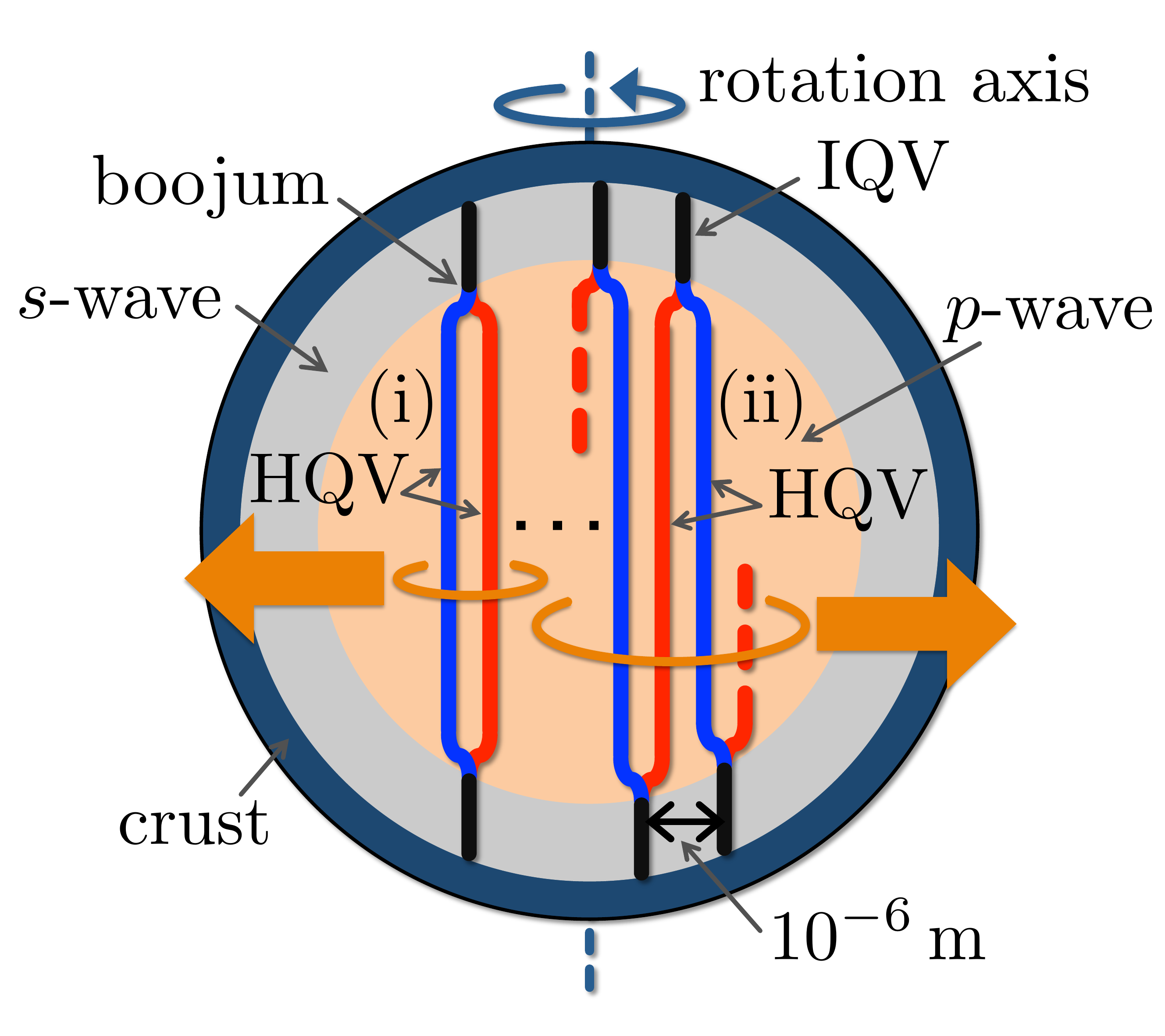}
\end{tabular}
\end{center}
\caption{
Inner structures of NSs. 
(a) Pinned vortices in the $s$-wave superfluid in the core 
surrounded by the crust (dark blue) in the outermost region.
(b) Vortex network in 
the $p$-wave inner core (pink) surrounded by the $s$-wave outer core of the spherical shell (grey).
(a) Vortices form a lattice and are pinned at nuclei in the crust (see the inset).
(b) Vortices are not pinned at nuclei in the crust. 
 A single integer quantum vortex (IQV) in the $s$-wave outer core is split into two half-quantized vortices (HQVs) 
in the $p$-wave inner core. 
(i) The same pair of two HQVs
in the $p$-wave inner core are connected to a single IQV
in the upper and lower  $s$-wave regions,  
forming a cluster of a certain minimum size.
(ii) Two HQVs connected to a single vortex in the upper (lower) $s$-wave region are connected to different IQVs in the lower (upper) 
$s$-wave region, 
forming a cluster with a larger size.
For (a) and (b), the orange arrows denote that vortices are released 
when a NS decreases its rotation speed, in which case all vortices in the same cluster have to be released at the same time for the example presented in (b).
}
\label{fig:schematic0}
\end{figure}
The other key ingredient of our study is 
{\it quantum vortices} in superfluids.  
Superfluids are fluids possessing zero viscosity which can support states with persistent flows. 
In laboratory experiments, $^4$He atoms become a superfluid at low temperatures.
In the interiors of NSs, neutrons
form Cooper pairs and exhibit superfluidity \cite{Migdal:1960}  
(see ref.~\cite{Sedrakian:2018ydt} for a recent review) 
 [see Figure~\ref{fig:schematic0} (a)],
which is consistent with 
the thermal evolution of NSs, the long relaxation time after each glitch~\cite{Baym1969,Pines1972,Takatsuka:1988kx}, and 
the neutrino emissivity 
\cite{Page:2010aw}.
When superfluids undergo rotation, vortices are created. 
One of remarkable properties of superfluids is the fact that 
these vortices 
carry a quantized amount of angular momentum unlike classical fluids,
hence the name {\it quantum vortices}. 
Since neutron stars are rapidly rotating, the superfluid component is pierced by quantum vortices lying along the rotation axis, which typically form an Abrikosov triangular lattice whose number can reach  $10^{19}$. 
There is also an accompanying ordinary component 
(not forming Cooper pairs) 
which lowers its rotation speed by releasing gravitational waves and electromagnetic pulses, 
whose period is observed to be a continuous quantity. 
On the other hand, the superfluid component maintains a constant rotation speed,  
as a consequence of superfluidity. Therefore, a gap between the rotation speeds of the two components grows over time. 
The superfluid component can lower its rotation speed by releasing vortices. If the vortices can be released one by one, then the superfluid component's rotation speed can catch up with the ordinary component {\it immediately} once this difference reaches the amount corresponding to one quantized vortex, and the change of the rotation should be (almost) 
smooth at any given time, which does not explain the appearance of glitches.

In order to overcome this problem, 
the hypothesis of catastrophic unpinning of quantum vortices \cite{Anderson:1975zze,Alpar1977}  
assumes that all vortices are pinned to nuclei, 
which play the role of impurities in the crust [see the inset of Figure~\ref{fig:schematic0} (a)],  
analogously to metallic superconductors where 
Abrikosov vortices are energetically favored to be pinned to                                                                                        
impurities. 
Then, in order to explain the occurrence of glitches, 
an avalanche of unpinning of a large number of vortices is assumed to occur 
spontaneously. 
Several models have been suggested, 
but one usually needs some phenomenological parameters 
 in order to account for the momentum transfer from the core region to the crust, 
a subject around which there is quite some uncertainty, fueling a long debate 
\cite{Alpar1984,
haensel2007neutron,
10.1093/mnras/sts108,
PhysRevLett.109.241103,PhysRevLett.110.011101,PhysRevC.90.015803}. 
A more serious drawback 
comes from
a recent work based on microscopic calculations~\cite{PhysRevLett.117.232701} showing that, unlike in metallic superconductors, 
pinning of vortices to impurities is energetically disfavored 
in the case of nuclear superfluids.

Here, we first reanalyse the accumulated up-to-date observation data~\cite{glitch-database,pulsar-database} 
to obtain 
the scaling law $P(E) \sim E^{-\alpha}$ 
with the exponent $\alpha \approx 0.88$. 
We then propose a simple microscopic model that
naturally deduces the scaling law with this exponent 
without any additional free parameters.
Our model explains the origin of the glitches in terms of 
a {\it quantum vortex network} 
that arises at the interfaces of the two different kinds of superfluids 
in the cores of NSs.
In contrast to the catastrophic vortex unpinning hypothesis, we need no (un)pinning for the accumulation of quantum vortices.

The two different kinds of superfluids 
explained above were theoretically predicted as two different types of neutron Cooper pairs: $s$-wave and $p$-wave parings. 
While an $s$-wave paring \cite{Migdal:1960} is dominant 
  in the low-density regime relevant for the neutron star outer core, 
a $p$-wave 
paring is dominant in the high-density regime relevant for the inner core 
\cite{Tabakin:1968zz,Hoffberg:1970vqj,Tamagaki1970,Hoffberg:1970vqj,Takatsuka1971,Takatsuka1972,
Richardson:1972xn,
saulsPRD78,muzikarPRD80,saulsPRD82,Yasui:2019unp}.\footnote{
More precisely, $s$-wave and $p$-wave pairings denote 
$^1S_0$ (spin-singlet and $s$-wave 
with total angular momentum $J=0$)
 and $^{3}P_{2}$ 
 (a spin-triplet and  $p$-wave with total angular momentum $J=2$) 
 parings, 
 respectively. 
 See Supplementary Information for more details.
 }
Therefore, we will assume that the interior of neutron stars consists of a layer structure with a $p$-wave inner core surrounded by an $s$-wave outer core forming a spherical shell 
[Figure~\ref{fig:schematic0} (b)].

\begin{figure}[ht]
\begin{tabular}{ccc}
(a) &(b) & (c)\\
\begin{minipage}{0.35\hsize}
   \vspace{-18em}
   \centering
   \includegraphics[scale=0.22]{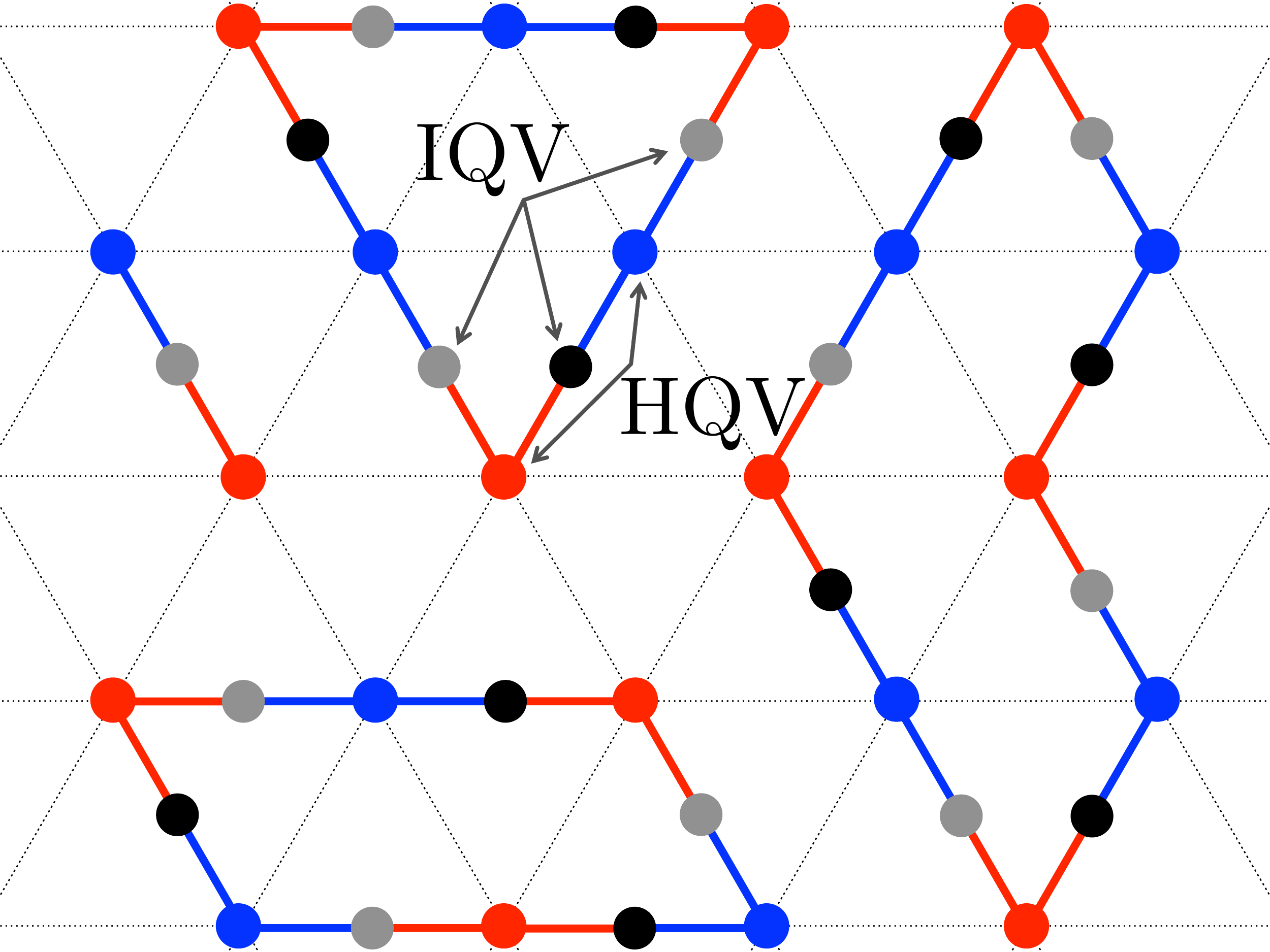}
\end{minipage}
&
\includegraphics[scale=0.20]{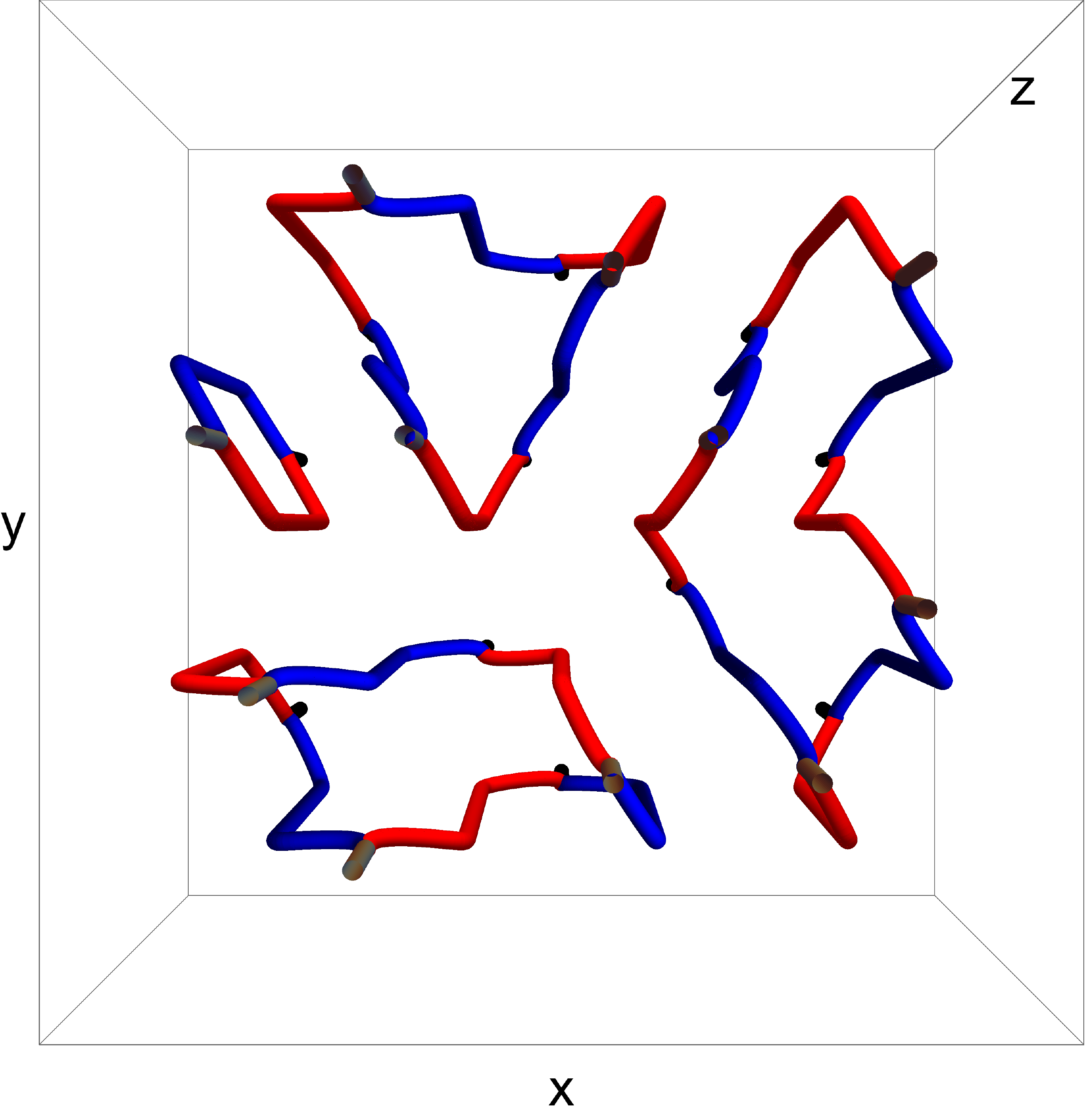} &
\includegraphics[scale=0.27]{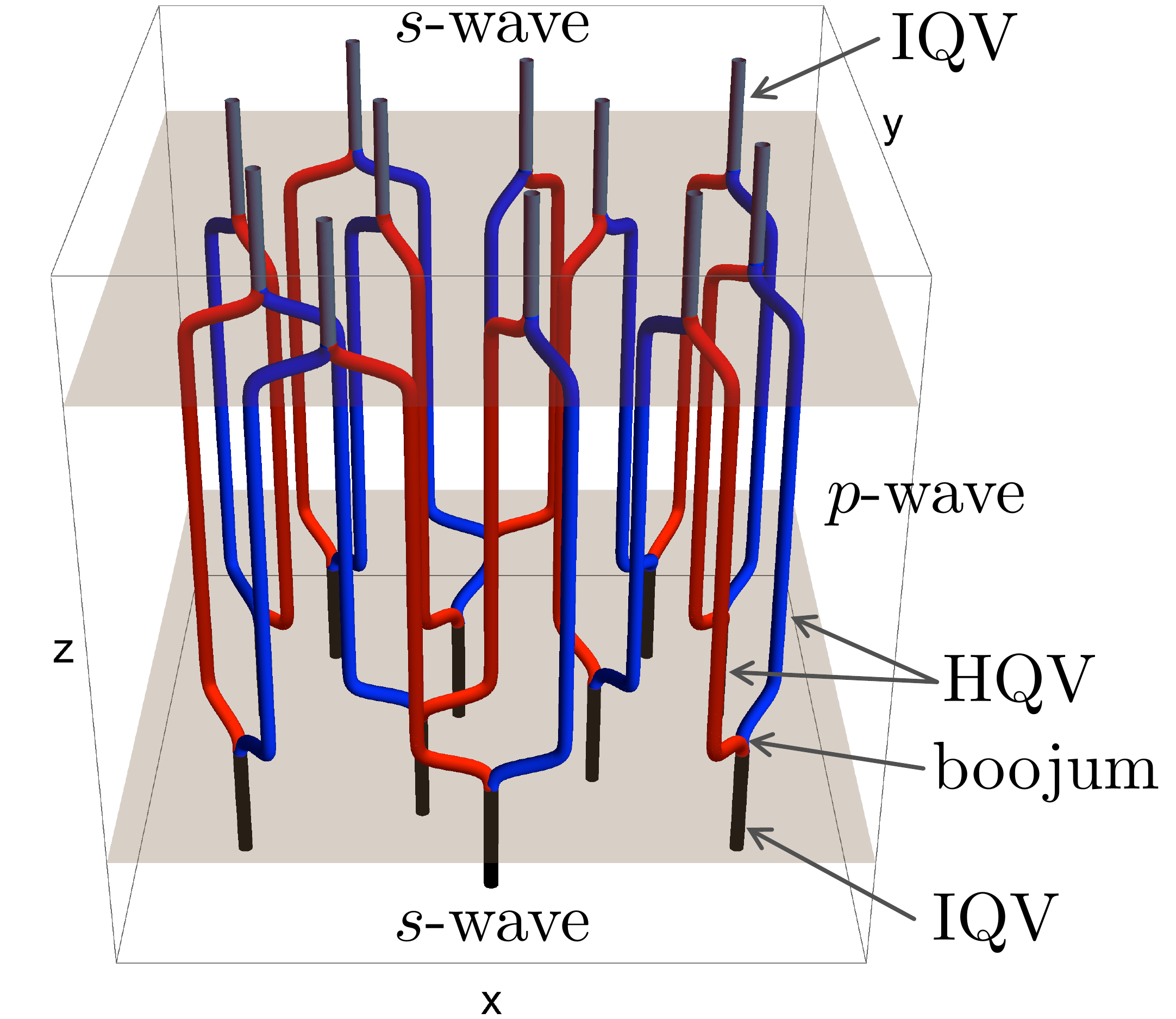}
\end{tabular}
\caption{Schematic views of pairings of vortices at $s$-$p$-$s$ interfaces resulting in vortex clusters. 
(a) top view of a vortex network, 
(b) top view 
and 
(c) view from diagonally above 
of a 3D configuration.
In (a), black and grey dots denote integer vortices in the $s$-wave region at the top and bottom, respectively, while red and blue dots and lines  denote 
HQVs in the $p$-wave region forming an Abrikosov lattice. 
 Clusters with sizes one, three and four are shown. 
}
\label{fig:schematic}
\end{figure}
We show that the glitch mechanism can be explained by the crucial property of $p$-wave superfluids:
 the existence of {\it half-quantized} vortices (HQVs)  
 carrying half-quantized circulations~\cite{Masuda:2016vak}, 
 as opposed to integer-quantized vortices (IQVs)~\cite{Richardson:1972xn,muzikarPRD80,saulsPRD82,masudaPRC16,Masaki:2019rsz}.
HQVs are energetically favored 
so that one IQV is split into two HQVs 
[denoted as blue and red vortices in Figure~\ref{fig:schematic0} (b)]
with additional topological charges cancelling each other (see Supplementary Information).
Quantum vortices in the $s$-wave and $p$-wave superfluids 
  are connected through junctions called ``boojums'' \cite{Mermin1977} at the interface;
{\it one} IQV in the $s$-wave superfluid is connected to {\it two} HQVs in the $p$-wave superfluid. 
We thus have a picture of a large number of vortices penetrating 
 the $p$-wave inner core surrounded by the $s$-wave outer core
 as in Figure~\ref{fig:schematic0} (b).
Then, it is possible that the same pair of two HQVs
in the $p$-wave core are connected to a single IQV 
in the upper and lower $s$-wave regions 
[see (i) in Figure~\ref{fig:schematic0} (b)].
If this is the case,
a neutron star superfluid can change its rotation speed by releasing vortices one by one from the core 
as is the case without vortex pinning at the crust, 
which would not explain the appearance of glitches. 
However, more generally, as illustrated in (ii) in Figure~\ref{fig:schematic0} (b), 
we can have two HQVs connected to a single IQV in the upper (lower) $s$-wave region 
which are connected to a different IQV in the lower (upper) 
$s$-wave region,
thereby leading to the formation of a vortex network composed of a cluster of connected vortices 
as in Figure~\ref{fig:schematic}. 
As we show below, in this case, each cluster can contain a large number of vortices, 
which exhibits the power-law distribution of glitches. 

\medskip
\begin{figure}[htb]
\includegraphics[scale=0.3]{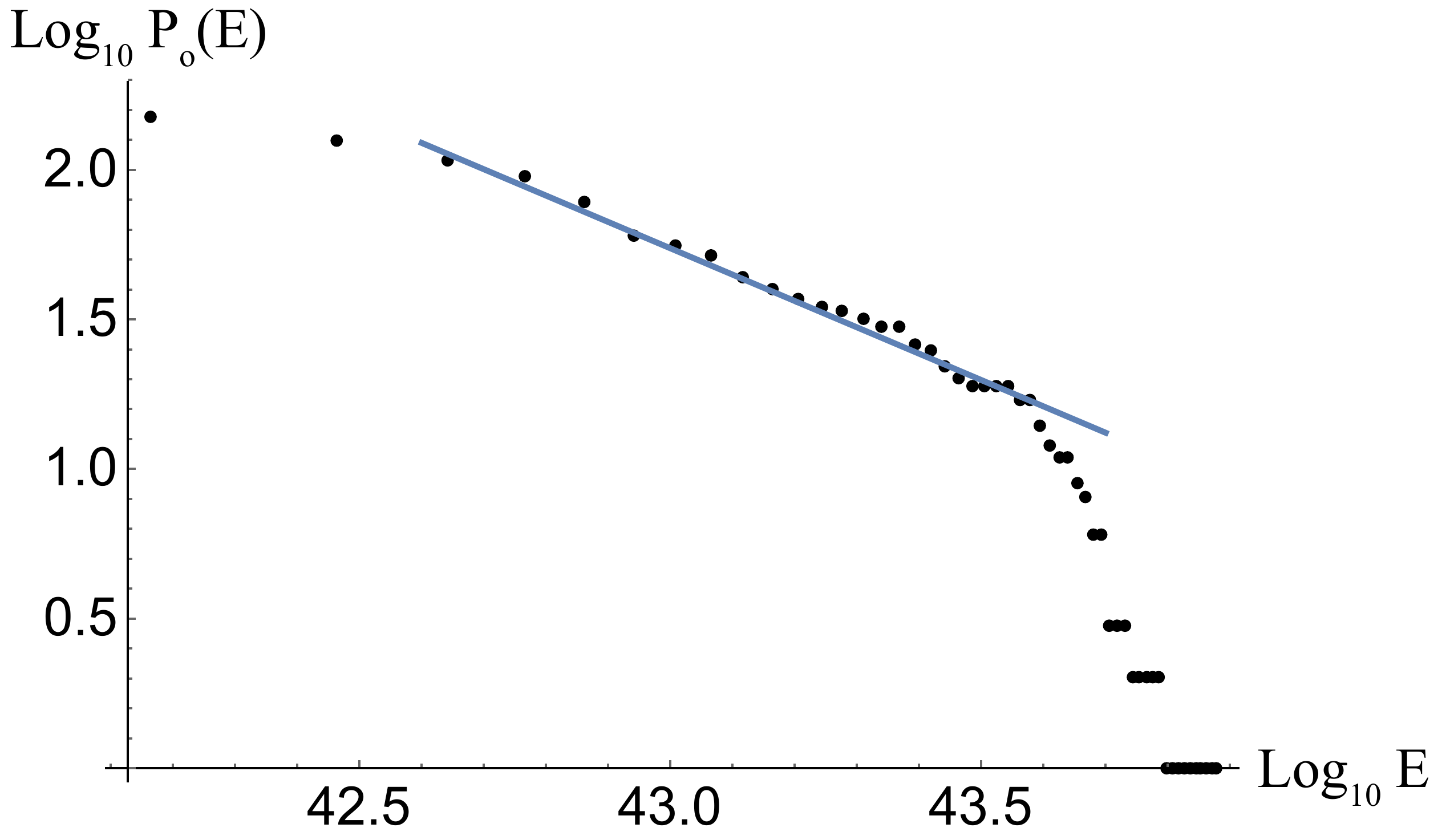}
\caption{$\log P(E)$ vs. $\log E$ plot of observed glitches of energy $E$.}\label{fig:logNlogE}
\end{figure}
\textit{Observed distribution of energy. -}
Let us analyze the glitch dataset reported by ref.~\cite{glitch-database}; we additionally use the pulsar catalogue~\cite{pulsar-database} to retrieve the pulsar periods. 
Let $P_o(E)$ denote the cumulative probability of observed glitch energies as in~\cite{morley-93} (see 
Appendix for details). 
Figure~\ref{fig:logNlogE} displays the log-log plot of $P_o(E)$ obtained from the analysis of 533 glitches,
where the line, determined by 
the least squares method using all 533 data points, 
shows  
the cumulative distribution of the power law 
\begin{equation}
P_o(E)\sim E^{-0.88\pm0.03}. \label{eq:scaling}
\end{equation}
One can see that in the central region (away from extremely small or extremely large glitches), the cumulative distribution is well approximated by the power law.

\medskip
\textit{Cluster size and energy distribution. -}
The radius of the whole core is typically 10km.
There is an uncertainty 
for the size of the $p$-wave inner core 
but it is thought to be a few km. 
The mean intervortex distance is of order $10^{-6}$m, and
the number of vortices is of order $10^{19}$.
We assume that the $s$-$p$ interfaces are mostly flat and parallel for simplicity although the outer and inner cores would be almost spherical. 
However, our results do not depend on the precise details of the underlying shapes and sizes at all.
We further assume a triangular vortex lattice 
which is rigidly rotating, as is typically the case for superfluids.
We consider a rotating frame in which this lattice is static.
Then, 
at the  $s$-$p$ interface, two HQVs with different topological charges 
in the $p$-wave region
will pair and connect to an integer vortex in the $s$-wave region via a boojum.

\begin{figure}[ht]
\centering
\begin{tabular}{ccc}
(a) & (b) & (c) \\
\includegraphics[scale=0.22]{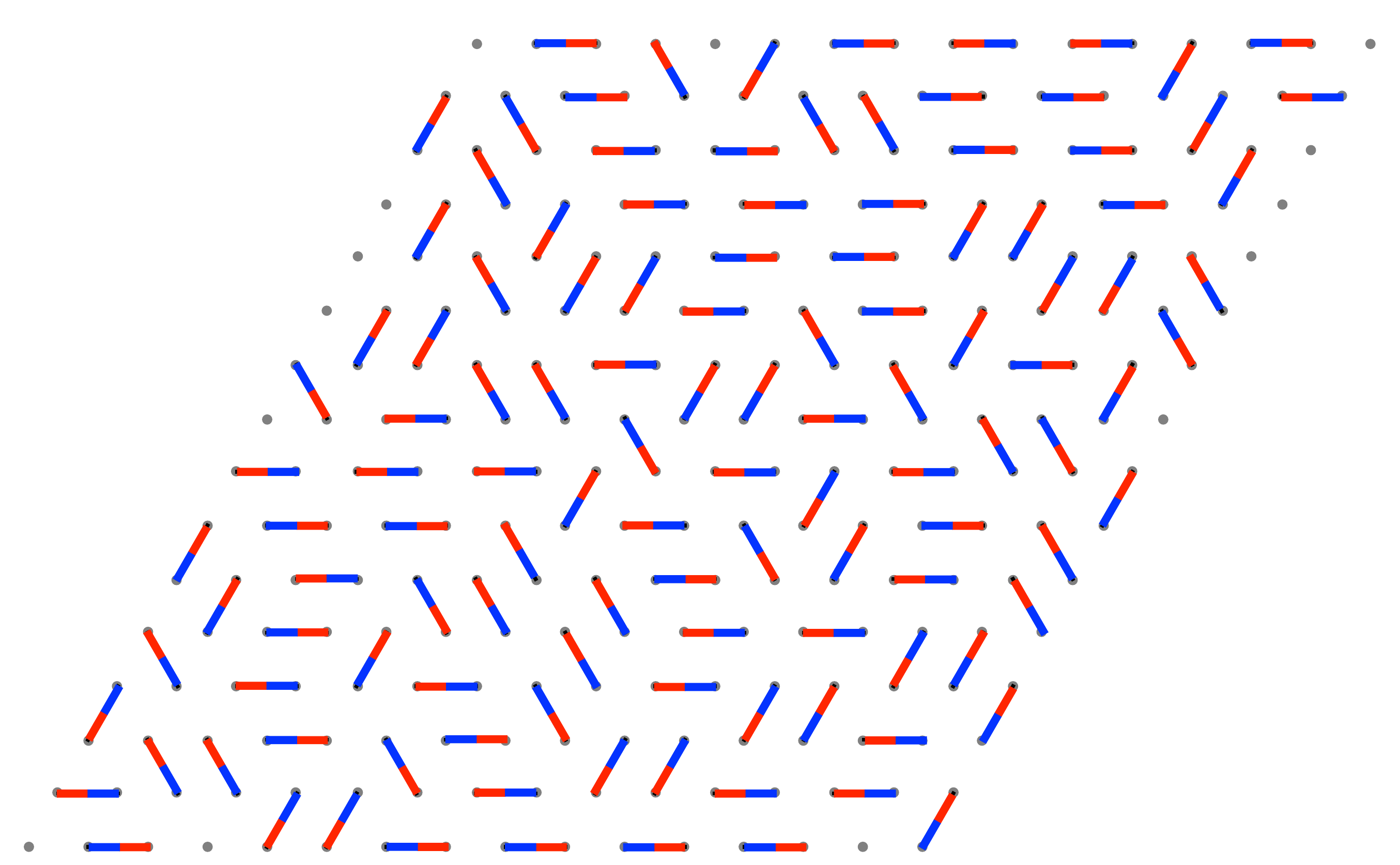} &
\hspace{-0.3cm}
\includegraphics[scale=0.22]{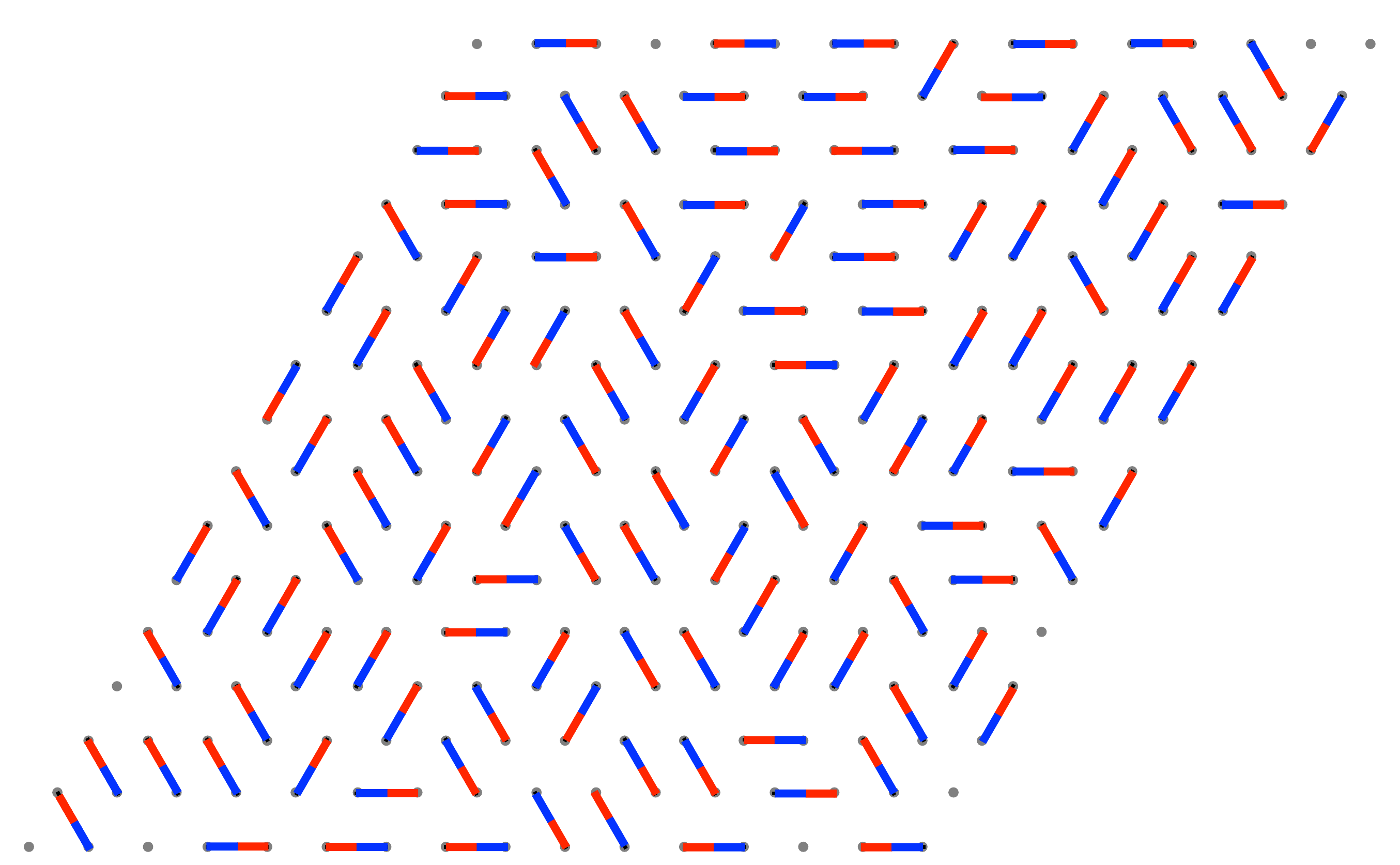} &
\hspace{-0.3cm}
\includegraphics[scale=0.22]{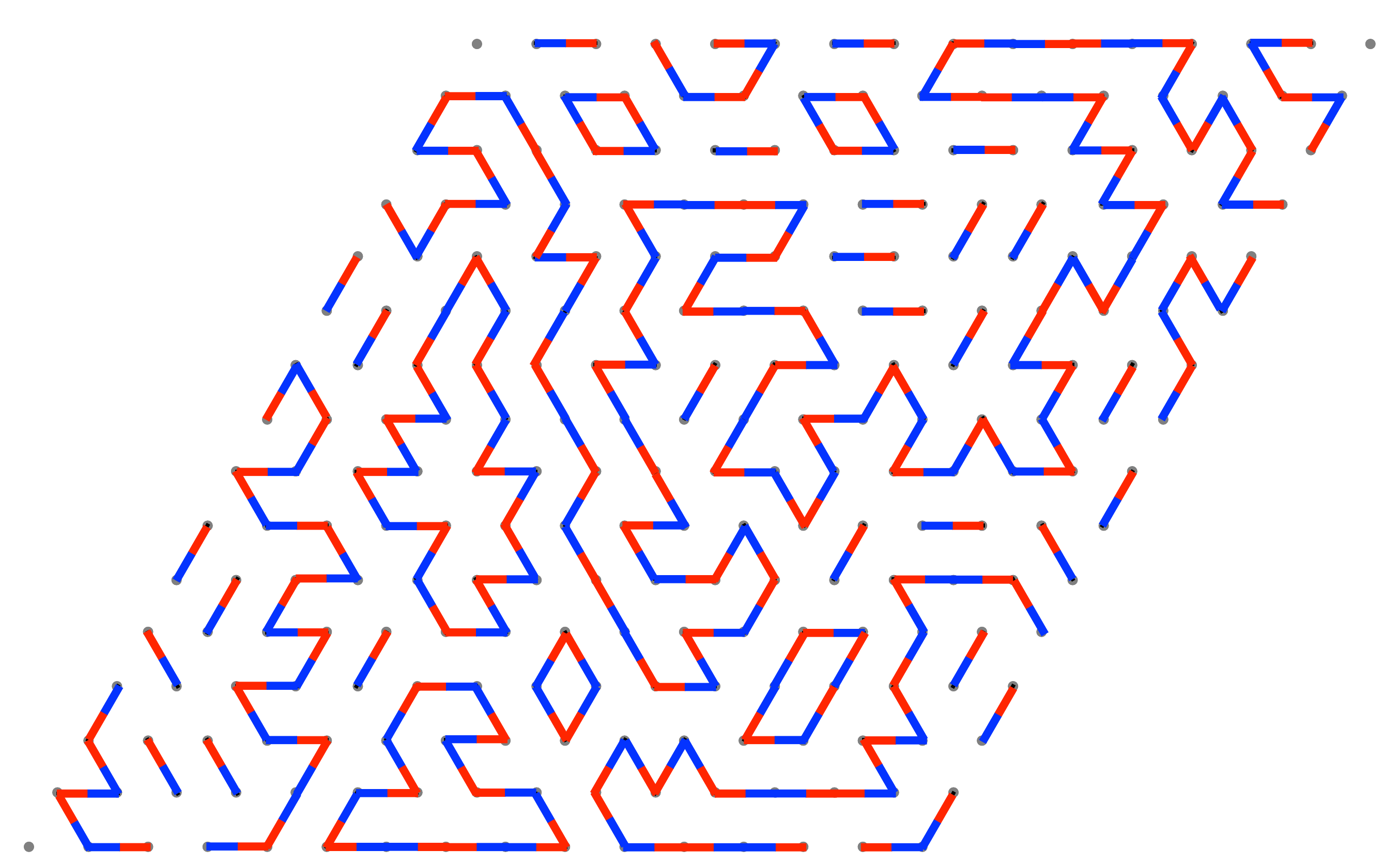}
\end{tabular}
\caption{Snapshot of pairings at the top (a) and bottom (b) interface, and the resulting vortex clusters (c), which appears as loops in the top view.}
\label{fig1}
\centering
\includegraphics[scale=0.25]{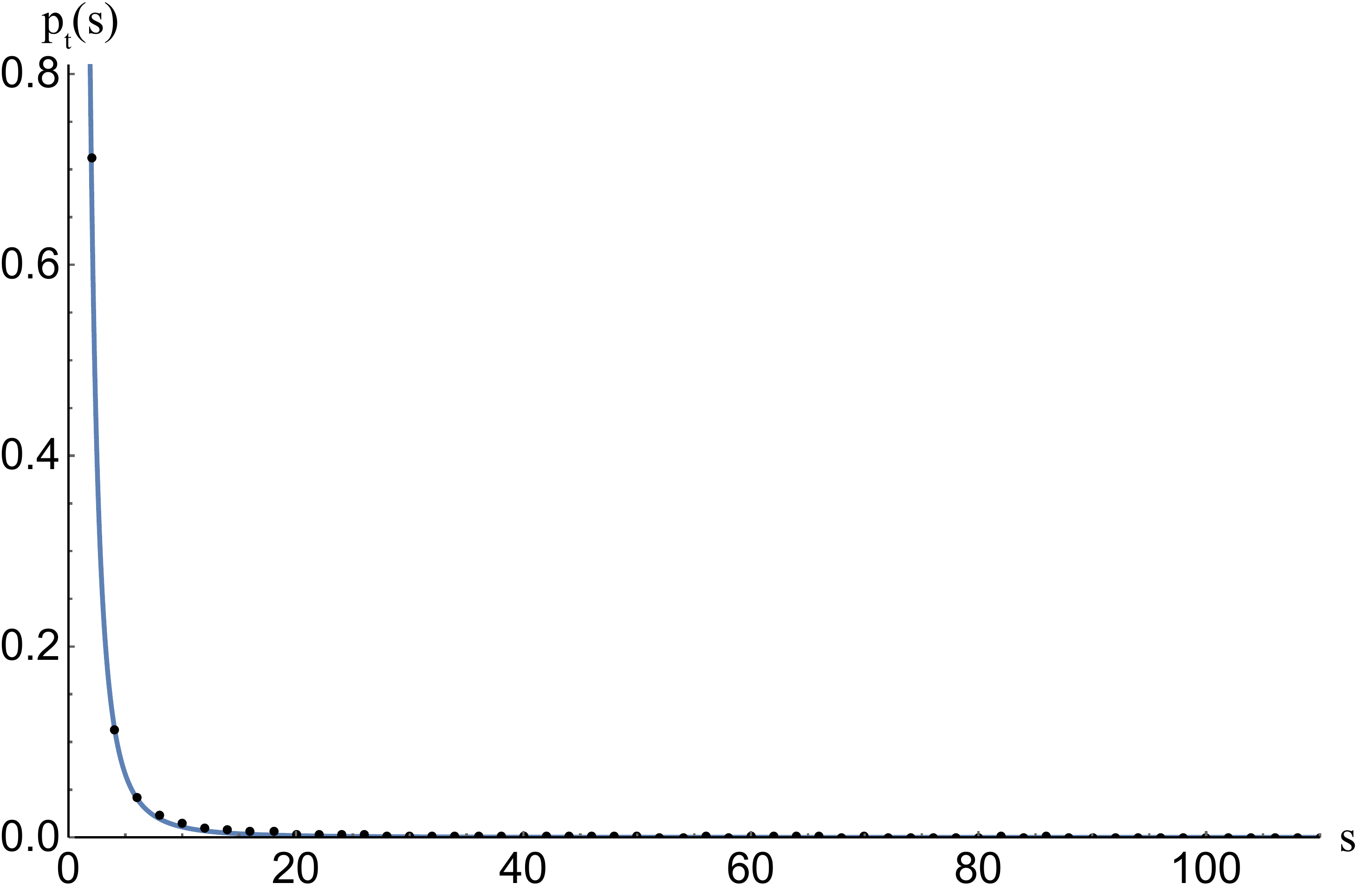} 
\caption{Plot in linear 
scale of the simulated cluster-size probability data points $\{s_i,q_i=p(s_i)\}$ and the inferred power law $p_t(s) \sim s^{-2.6 }$ (see 
Appendix for details).}
\label{fig:bayes0}
\end{figure}
In order to study the cluster size distribution we employ the following model. We assume that the HQVs in the $p$-wave region form an Abrikosov-like triangular lattice with periodic boundary conditions; however we do not assume any restriction on the blue/red vortex pattern except that for a red vortex it is always possible to find a blue nearest neighbor and vice versa. We then consider 
a triangular vortex lattice of  rhombic shape 
and simulate two random blue-red pairings, that would correspond to the boojums at the top and bottom ends of the vortices. The superposition of the two pairings generates loops of various sizes, which corresponds to clusters in which all vortices are connected via boojums. 
This was schematically depicted 
in Figure~\ref{fig:schematic},
and an example of configurations taken from our simulation 
is presented in Figure~\ref{fig1}
 (note that a minimal loop made of two vortices appears in the form of a segment).

Statistical analysis of the cluster size distribution shows a power-law behaviour (for details of the analysis see 
Appendix). The best estimate of the probability distribution of cluster size is 
\begin{equation}
p_t(s) \sim  s^{-2.6\pm 0.3}, \label{eq:pts}
\end{equation}
where $s$ is the cluster size 
and the subscript $t$ stands for ``theoretical''; 
the simulated data points and the best fit to Eq.~(\ref{eq:pts}) is displayed in Figure~\ref{fig:bayes0}. 
Then, we can define the cumulative probability as
$P_t(s) = \int_s^{s_{max}} p_t(u) du$. 
A cluster of size $s$ defines a region inside of which the number of vortices is of order $s^2$. Since there is no reconnection between HQVs with different topological charges \cite{Mermin:1979zz,Kobayashi:2008pk}, when a cluster is expelled from the neutron star core, it necessarily drags all the other vortices enclosed by that cluster. It is therefore safe to 
assume that the energy associated with the emission of a vortex cluster satisfies the relation $E=c s^2$ for some constant $c$.  
By using this relation, we can translate the size distribution in Eq.~(\ref{eq:pts}) to the energy probability distribution, $p_t(E) \sim E^{-1.8\pm 0.2}$  and the corresponding cumulative distribution (see 
Appendix) 
\begin{eqnarray}
 P_t(E) \sim E^{-0.8\pm 0.2}.
\label{eq:scaling2}
\end{eqnarray}
Thus, our model  gives a description of the scaling law in 
Eq.~(\ref{eq:scaling}) 
for the set of observed glitches without any free parameters.

\medskip
\textit{Conclusion.-}
 We have obtained the power-law scaling law of Eq.~(\ref{eq:scaling}) 
 for glitches from recent observational data,
and proposed a simple model to explain the scaling law of Eq.~(\ref{eq:scaling2}) based on vortices penetrating 
the $p$-wave superfluid core surrounded by 
the $s$-wave superfluid. 
 Boojums connecting two HQVs in the $p$-wave superfluid 
and one integer vortex in the $s$-wave superfluid give rise to clusters of vortices,
whose distributions realise the scaling law. 
The appearance of the vortex network naturally explains 
a similarity with power-law distributions in other systems widely discussed in  network science.
The key strength of our approach is that 
our model contains 
 no free parameters 
to explain the observed data. 

\subsection*{Acknowledgements} 
We thank Matthew Edmonds, 
Antonino Flachi, Daisuke Inotani, 
Hiromichi Kobayashi, Calum Ross  
and Norihiko Sugimoto for reading the manuscript and giving us useful comments, 
and Ryosuke Yoshii for a discussion at the early stage of this work.
This work is supported by the Ministry of Education, Culture, Sports, Science (MEXT)-Supported Program for the Strategic Research Foundation at Private Universities ``Topological Science" (Grant No. S1511006). 
This work is also supported in part by 
JSPS Grant-in-Aid for Scientific Research [KAKENHI Grants No.~17K05435 (S.Y.), 
No.~16H03984 (M.N.), and No.~18H01217 (M.N.)], 
and also by MEXT KAKENHI Grant-in-Aid for Scientific Research on Innovative Areas ``Topological Materials Science'' No.~15H05855 (M.N.).

\begin{appendix}
\section*{Appendix}

\subsection*{Observed distribution of glitch energy} 
Let us analyze the glitch dataset reported by ref.~\cite{glitch-database}; we also use the pulsar catalogue~\cite{pulsar-database} to retrieve the pulsar periods.
The energy of a glitch can be estimated from the change in rotational energy of the neutron star, namely $E_g= I\,\Omega\, \Delta \Omega$, where $I$ is the moment of inertia and $\Omega$ is the rotational frequency. By taking $1.4M_\odot$ as the typical neutron star mass and $10^5$ cm as the typical radius, the glitch energy becomes~\cite{morley-93}
\begin{equation}
E_g \approx \frac{a\cdot 10^{46}}{P^2} \frac{\Delta \nu}{\nu} \mathrm{erg},
\label{eq:morley}
\end{equation}
where $P$ is the period of the pulsar in seconds, $\Delta \nu / \nu$ is the relative change of frequency associated with the glitch and $a$ is a constant of order one (for concreteness we take $a=2.91$~\cite{morley-93}).
Let us also define the cumulative probability of observed glitch energies as in~\cite{morley-93}:
\begin{equation}
P_o(E) = \int_E^{E_{max}} p_o(u) du.
\end{equation}

Figure~\ref{fig:logNlogE} displays the log-log plot of $P_o(E)$ obtained from the analysis of 533 glitches,
where the line, determined by 
the least squares method using all 533 data points, 
shows  
the cumulative distribution of the power law in Eq.~(\ref{eq:scaling}).
Note that a change of the overall energy scale in Eq.~(\ref{eq:morley}) would not affect this behaviour.

This scaling law is significantly different from the previously found scaling behaviour
$P_o(E) \sim E^{-0.14}$ \cite{morley-93}. 
The reason can be considered as follows:
ref.~\cite{morley-93} used 27 data points of glitches 
in the energy range  $10^{39}$ -- $10^{42}$ erg, 
for which the upper limit 
came from the observational limitations at that time, 
while our analysis with 
 533 data sets covers a broader range of energies than theirs 
 per Figure~\ref{fig:logNlogE}, 
 which yields the observed differences of the exponents.
Another important point is that several datum corresponding to  
extremely giant glitches with energy higher than $10^{43.6}$ erg 
are rare events, with corresponding lower probability, 
giving a smaller contribution to our fitting. 

\subsection*{Simulation of vortex pairing and clustering}

\begin{figure}[htb]
\centering
\begin{tabular}{cc}
(a) & (b) \\
\includegraphics[scale=0.27]{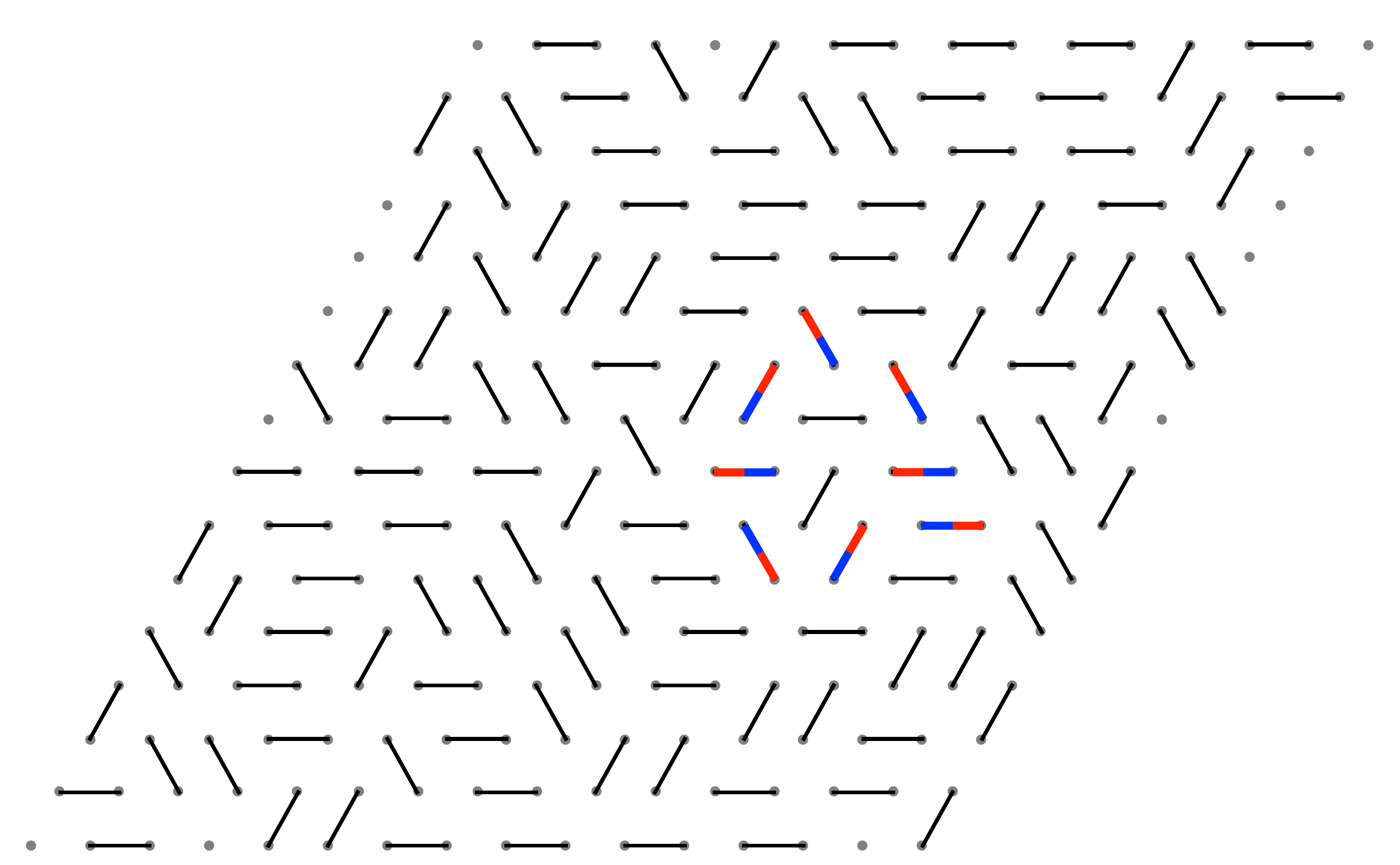} &
\includegraphics[scale=0.27]{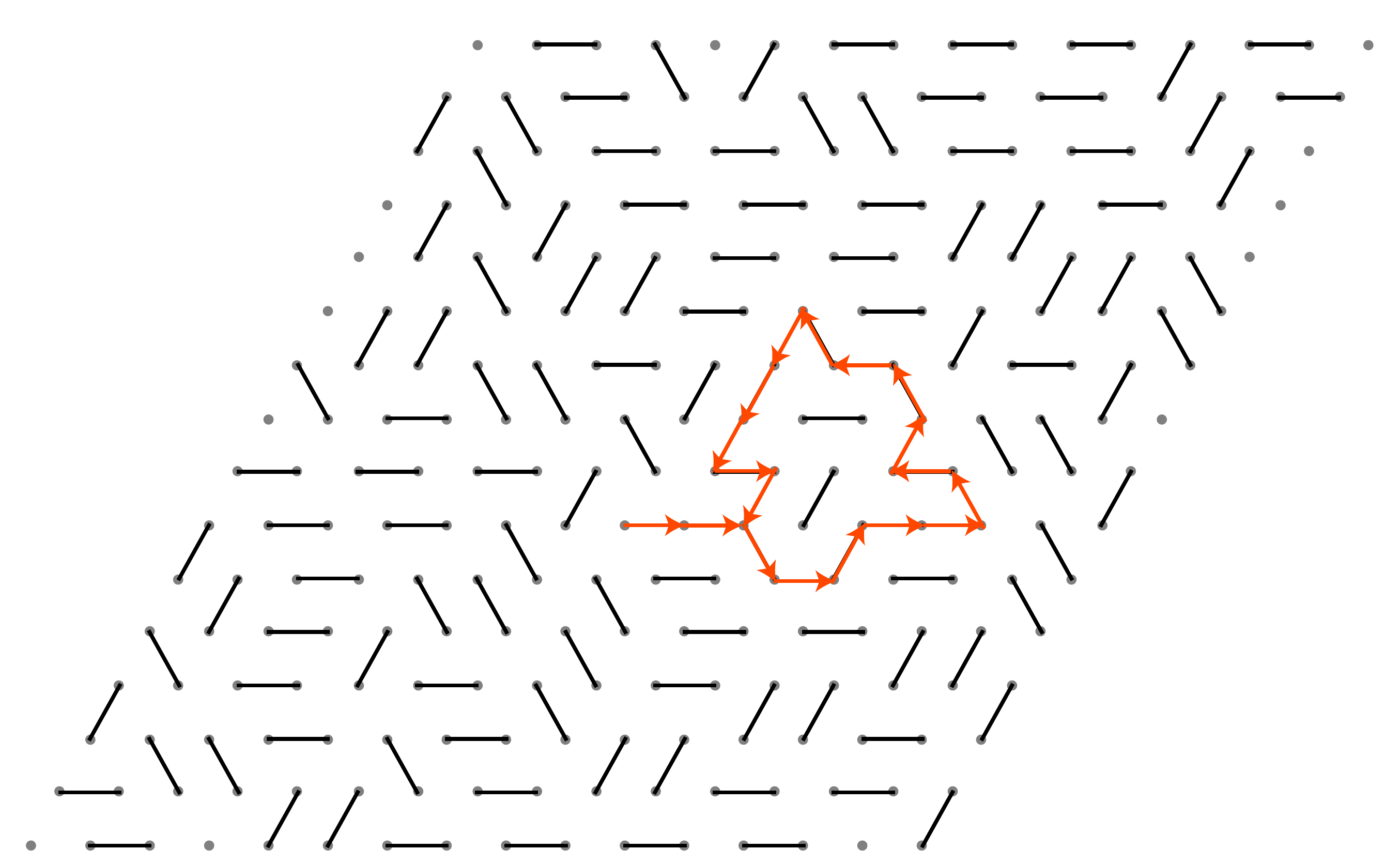} \\
(c) & (d) \\
\includegraphics[scale=0.27]{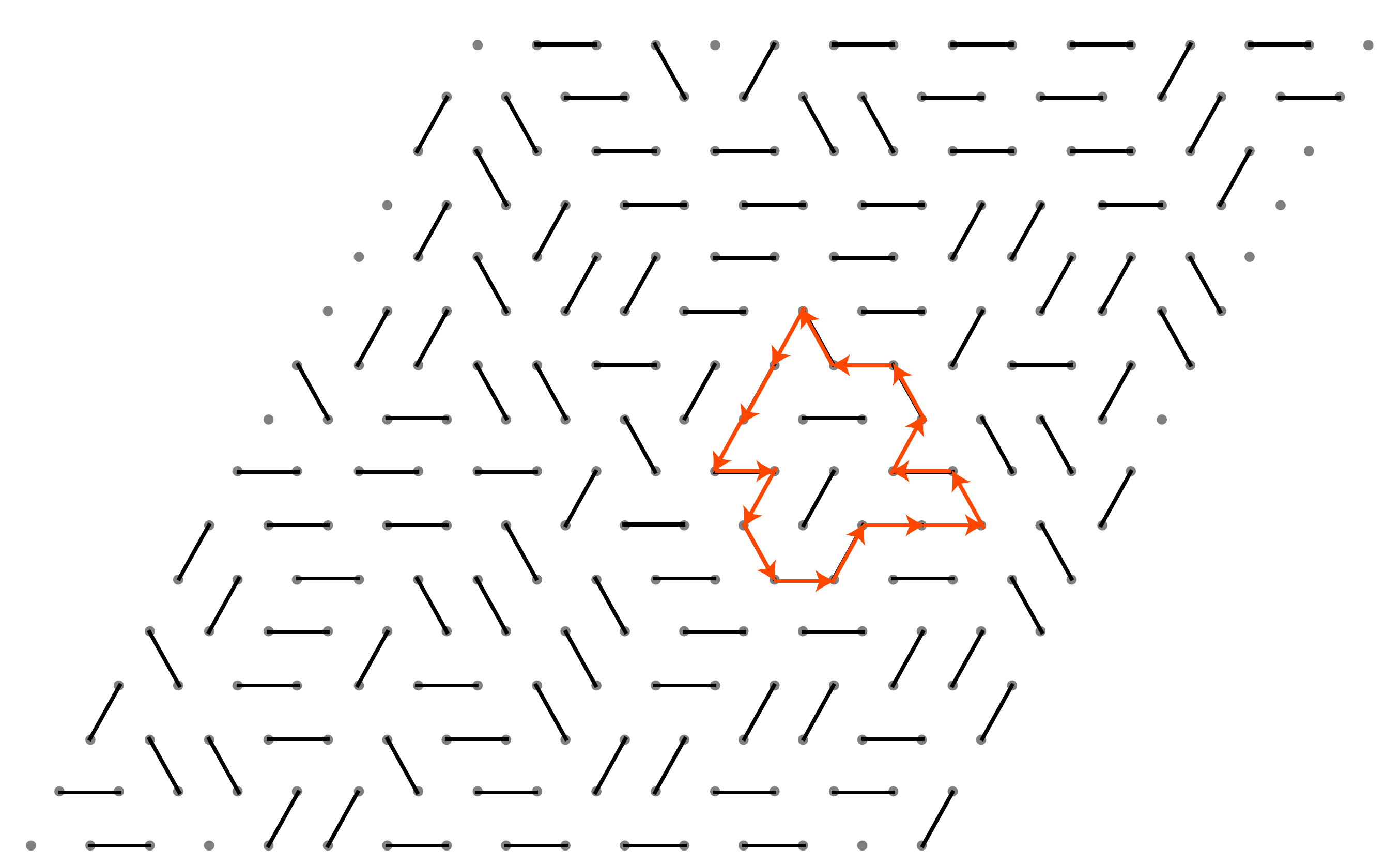} &
\includegraphics[scale=0.27]{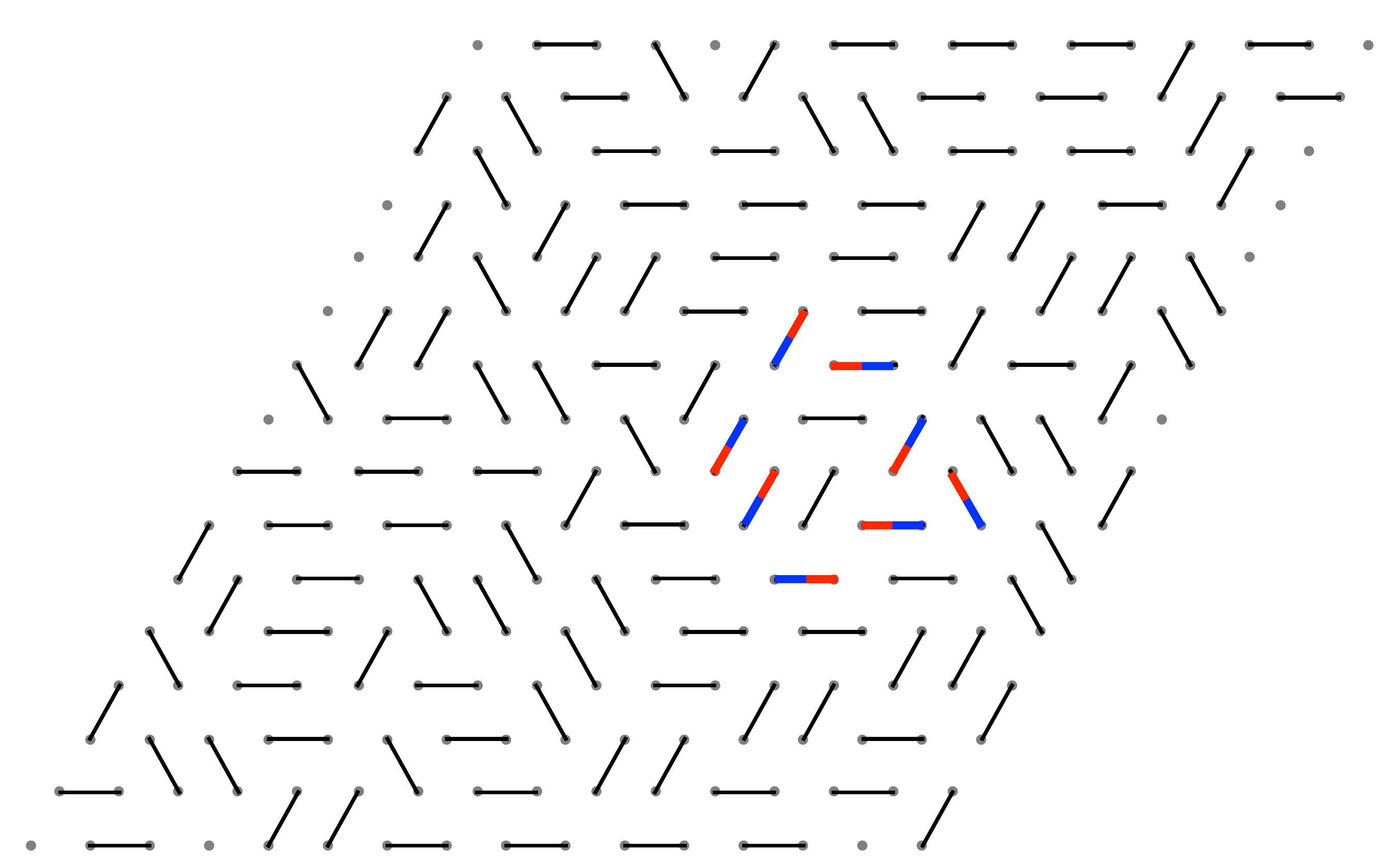}
\end{tabular}
\caption{Schematic representation of the worm algorithm: (a) starting configuration; (b) random generation of the worm, which terminates when it crosses itself; (c) only the loop section of the worm is retained; (d) pairs are flipped along the worm, giving rise to the new configuration. 
In (a) and (d), only pairs before and after flipping are coloured by red and blue.
}
\label{fig:worm}
\end{figure}
Cluster configurations are generated by simulating two uncorrelated random pairings 
in a triangular lattice with periodic boundary conditions.
 This is achieved by a version of the worm algorithm. First, we choose an arbitrary pairing, e.g. the simple pairing in Figure~\ref{fig:worm}, we then perform a series of decorrelation steps to obtain random pairings. Each step consists of the following operations. First we generate a path defined by a sequence of sites as follows: we randomly choose a site, that will be the first in the sequence; then we add the site to which the first one is paired; next, we randomly choose one of the five available nearest neighbours of the latter site (if it is a paired site already, which is already in the sequence, then it cannot be chosen) and add it to the sequence, followed by the corresponding paired site, and so on, until we encounter a site that is already in the sequence. This defines a path that contains a loop. If the loop has an odd number of segments, we discard it; otherwise we flip the pairing along the loop. This completes one decorrelation step; we perform $10^3$ steps for each site to obtain the first configuration (interpreted as the boojums
  at the top $s$-$p$-wave interface) and the same number of steps to obtain the second configuration (interpreted as the boojums at the bottom $s$-$p$ interface). Superposing the two configurations gives a schematic top view of the HQV system, in which loops represent vortex clusters that are kept together by boojums. The size of the clusters is 
calculated by a standard analysis of the adjacency matrix, which is the accepted procedure for complex networks.
See Supplementary Information (Sec.~c) for possible realisations in condensed matter setups of $s$-$p$-$s$ interfaces.

\subsection*{Statistical analysis}

\begin{figure}[htb]
\centering
\includegraphics[scale=0.25]{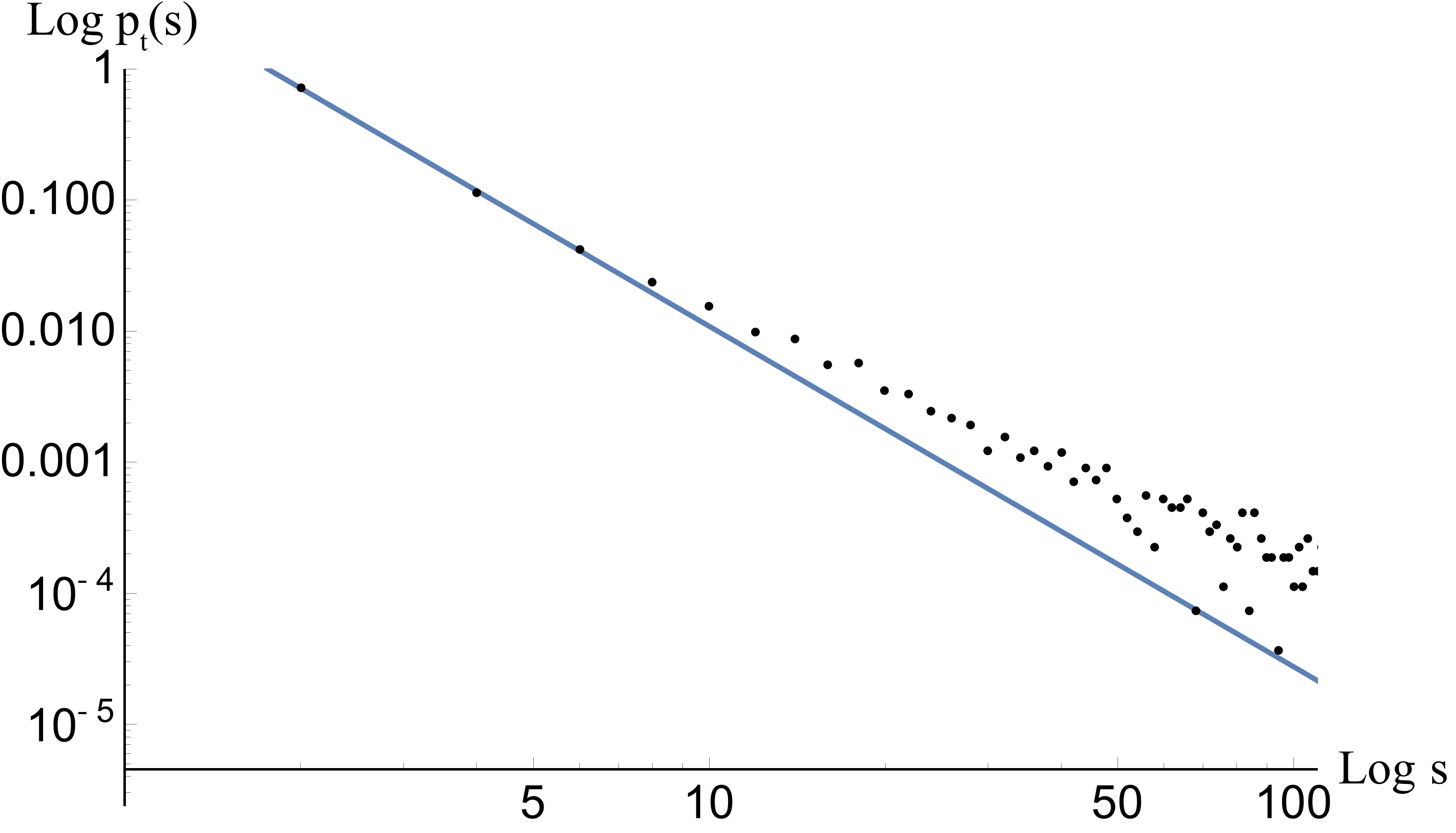} 
\caption{Plot in logarithmic  
scale of the simulated cluster-size probability data points $\{s_i,q_i=p(s_i)\}$ and the inferred power law $p_t(s) \sim s^{-2.6 }$.
}
\label{fig:bayes}
\end{figure}
In this section we statistically analyze 100 independent simulations of size with $48^2=2304$ sites. 
We use a Bayesian technique based on Dirichlet distributions~\cite{delpozzo-2018}, which is appropriate for discrete random variables, such as our cluster size. The likelihood of observing a data set is given by the multinomial distribution
\begin{equation}
p(n_1,\ldots, n_k|q_1,\ldots,q_k) = \frac{N!}{n_1!\ldots n_k!} \prod_i^k q_i^{n_k}, \label{eq:likelihood}
\end{equation}
where $n_i$ is the number of observed clusters with size $s_i=2 i$, $N=\sum_i n_i$ and $q_i$ are the corresponding probabilities. Since our simulations are independent, we simply consider the data from all simulations at the same time. We can infer the the probabilities $\mathbf{q}=\{q_i\}$ given the observed counts $\mathbf{n}=n_i$ using Bayes' theorem,
\begin{equation}
p(\mathbf{q} | \mathbf{n}) = \frac{p(\mathbf{n} | \mathbf{q}) \, p(\mathbf{q}) }{\displaystyle{\int} d\mathbf{q}\, p(\mathbf{n} | \mathbf{q})  \, p(\mathbf{q}) },
\end{equation}
where $p(\mathbf{n} | \mathbf{q}) $ is the likelihood defined in Eq.~(\ref{eq:likelihood}) and $p(\mathbf{q})$ is the prior  distribution of the probabilities $\mathbf{q}$. The appropriate choice of prior is the Dirichlet distribution 
\begin{equation}
\mathrm{Dir}(\mathbf{q} | \mathbf{a} ) = \frac{\Gamma(A)}{\prod_i^k  \Gamma(a_i) } \prod_i^k q_i^{a_i}, \qquad A\equiv \sum_i^k a_i
\end{equation}
with $a_i=1, \, \forall i$; this is chosen since we do not assume any specific knowledge about the probabilities before doing the simulations. For the properties of the Dirichlet and multinomial distribution the posterior distribution will also be of Dirichlet type, namely
\begin{equation}
p(\mathbf{q} | \mathbf{n}) = \mathrm{Dir}(\mathbf{q} | \mathbf{a} + \mathbf{n} ) . \label{eq:posterior}
\end{equation}
The expectation value of the probabilities will then be
\begin{equation}
\bar{q}_i = \frac{n_i+1}{N+k}.
\end{equation}
The error bars for the $q_i$'s can be estimated by taking the square root of the variance $\mathrm{Var}[q_i]= \mathrm{E}[(q_i-\bar{q}_i)^2]$ over the distribution Eq.~(\ref{eq:posterior}); they turn out to be quite small, essentially invisible in Figure~\ref{fig:bayes}.
We try to fit the data $\{s_i,q_i\}$ with two different models, namely exponential and power law, with a least-squares method, in order to understand the general behaviour of the cluster size distribution. By looking at the Akaike information criterion 
(see Supplementary Information (Sec.~d) for details), it is clear that that the power-law model is preferred and the best fit gives 
\begin{equation}
p(s) \sim s^{-2.6 \pm 0.3} \label{eq:psmeth}
\end{equation}
where $\pm 0.3$ denotes the statistical error.
In Figure~\ref{fig:bayes0} in the main text 
and in Figure~\ref{fig:bayes} 
the data $\{s_i,q_i\}$ and the inferred law of Eq.~(\ref{eq:psmeth}) are represented both in linear and 
logarithmic scales, respectively. 
Figure~\ref{fig:bayes} shows the log-log plot, in which the line is determined by the least square method using all 100 simulations. 
The large size configurations on the right part deviate from the line, 
but 
these deviations are quite tiny ($\approx 10^{-4}$--$10^{-3}$) 
giving a smaller contribution to our fitting.     
In fact, one can find that the fitting is very good in the linear plot in Figure~\ref{fig:bayes0}.  

We have also generated simulations with 
smaller $N$ and get similar results, 
thereby implying the $N$ independence of our results for large $N$.

\subsection*{Translation from the cluster size distribution to the glitch energy distribution}
By using the relation $E= c s^2$ between the glitch energy $E$ and the vortex cluster size $s$, 
we can translate the size distribution in Eq.~(\ref{eq:psmeth}) to the (cumurative) energy distribution of glitchs as
\begin{eqnarray}
P_t(s) &=& 4.8\int_s^{s_{max}} u^{-2.6} du 
= c' \int_{E=cs^2}^{E_{max}=cs^2_{max}} v^{-1.3} \frac{dv}{v^{1/2}} \nonumber \\
&=& c' \int_{E}^{E_{max}} v^{-1.8} dv = c'' E^{-0.8} = P_t(E) 
\end{eqnarray}
with some constants $c'$ and $c''$, 
which defines the energy probability distribution, $p_t(E) = c' E^{-1.8\pm 0.2}$  and the corresponding cumulative distribution 
\begin{eqnarray}
 P_t(E) = c'' E^{-0.8\pm 0.2}.
\label{eq:scaling2-}
\end{eqnarray}

%
%

\section*{Supplementary Information}

\subsubsection{Comments on $p$-wave paring 
and half-quantized vortices}

As mentioned in the main text, 
in the interiors of NSs, neutrons
form Cooper pairs and exhibit superfluidity \cite{Migdal:1960}  
(see refs.~\cite{Chamel2017,Haskell:2017lkl,Sedrakian:2018ydt} for recent reviews).
Migdal considered an $s$-wave ($^1S_0$: spin-singlet and $s$-wave 
with total angular momentum $J=0$) paring  \cite{Migdal:1960} as conventional ($s$-wave) metallic superconductors. 
Although the $^1S_0$ channel is attractive dominantly in the low-density regime,
it becomes repulsive in the high-density regime in the neutron star core.
There, the attraction is provided by the $^{3}P_{2}$ channel with the spin-orbit ($LS$) force 
inducing the neutron $^{3}P_{2}$ (spin-triplet and $p$-wave 
with total angular momentum $J=2$) superfluids~\cite{Tabakin:1968zz,Hoffberg:1970vqj,Tamagaki1970,Hoffberg:1970vqj,Takatsuka1971,Takatsuka1972,Fujita1972,Richardson:1972xn,
saulsPRD78,muzikarPRD80,saulsPRD82,
masudaPRC16,Masuda:2016vak,Masaki:2019rsz,Yasui:2019unp,
Amundsen:1984qc,Takatsuka:1992ga,Baldo:1992kzz,Elgaroy:1996hp,Khodel:1998hn,Baldo:1998ca,Khodel:2000qw,Zverev:2003ak,Maurizio:2014qsa,Bogner:2009bt,Srinivas:2016kir,
Chatterjee:2016gpm,
Mizushima:2016fbn,
Yasui:2018tcr,Yasui:2019pgb,Yasui:2019vci,Mizushima:2019spl,Yasui:2020xqb}. 
The ground state at the weak coupling limit is in the so-called nematic phase 
\cite{saulsPRD78,muzikarPRD80}. 
We also comment that, although the boundary between the $s$- and $p$-wave regions may not be sharp microscopically 
(in reality an overlap region is predicted to exist 
\cite{Yasui:2020xqb}),  from a macroscopic point of view one can assume the presence of such an interface. 
 
The 
$p$-wave pairings are widely considered in condensed matter physics 
such as $p$-wave superconductors 
for which electrons form Cooper pairs 
and $^3$He superfluids for which $^3$He atoms form Cooper pairs.
Recently, $p$-wave superconductors (superfluids) have attracted 
great interests as
topological superconductors (superfluids) 
\cite{PhysRevB.78.195125,Kitaev2009}, 
allowing topologically protected Majorana fermions on the boundary surfaces and 
vortex cores.
In fact, the $^3P_2$ neutron superfluids 
have also been shown to be topological superfluids 
admitting surface Majorana fermions 
\cite{Mizushima:2016fbn}  and
Majorana bound states in their vortex cores  
\cite{Masaki:2019rsz}.
Thus, neutron star cores may be 
the largest topological materials in our universe.

The order parameter for $p$-wave  paring 
(for $^3P_2$ neutron superfluids) 
is described by a $3 \times 3$ traceless symmetric matrix $A$ with complex components \cite{saulsPRD78,muzikarPRD80}. 
Then, vortices take the following form:
an IQV is represented as
\begin{eqnarray}
 A = e^{i \theta} A_0
\end{eqnarray}
  with a spatial angle $\theta$ ($0 \leq \theta < 2\pi$) around the vortex core 
\cite{Richardson:1972xn,muzikarPRD80,saulsPRD82,masudaPRC16}, 
where $A_0$ is a constant matrix representing the ground state configuration. 
On the other hand, a HQV takes the form of 
\begin{eqnarray}
 A = e^{i {\theta \over 2}} O(\theta) A_0 O^T(\theta) , 
\quad 
  O(\theta) = \left(\begin{array}{ccc}\cos {\theta \over 4} & \pm \sin {\theta \over 4} & 0\\
 \mp \sin {\theta \over 4} & \cos {\theta \over 4} & 0 \\
 0 & 0& 1\end{array}\right), 
\end{eqnarray}
with the $D_4$ biaxial nematic ground state $A_0 \sim {\rm diag.}(1,-1,0)$,  
 where the sign $\pm$ ($\mp$) corresponds to HQVs with different topological charges cancelling each other \cite{Masuda:2016vak}, 
 denoted by red and blue in the main text.
The exponential factors 
$e^{i \theta}$ and $e^{i {\theta\over 2}}$ for IQV and HQV are  
the origin of  integer and half quantizations, respectively.

\subsubsection{A story of the boojum}
  Originally the boojum is a particular variety of  
  the fictional animal species called {\it snarks} created by Lewis Carroll in his nonsense poem ``The Hunting of the Snark.'' 
  The similar structures found in helium superfluids were named boojums by Mermin~\cite{Mermin1977}.
  See~\cite{Mermin1990} for the story  how he made ``boojum" an internationally accepted scientific term.
  Now boojums have been predicted to occur in $^{3}$He superfluids~\cite{volovik2003universe} 
  in particular at the A-B phase boundary~\cite{PhysRevLett.89.155301,Bradley2008}, 
  liquid crystals~\cite{Carlson:1988}, 
  Bose-Einstein condensates~\cite{Kasamatsu:2013lda},  
  quantum field theory~\cite{Gauntlett:2000de,Shifman:2002jm, 
  Isozumi:2004vg}
  and high density quark matter relevant for neutron star cores~\cite{
  Cipriani:2012hr,
  Chatterjee:2018nxe}.

\subsubsection{Proposals for laboratory experiments}
We further point out that our mechanism 
for glitches by vortex networks through boojums can be  
simulated by laboratory experiments.
One is with ultracold atomic gases, 
a mixture of 
a scalar Bose-Einstein condensate (BEC) 
and a spin-2 nematic BEC 
\cite{Song:2007ca,Uchino:2010pf,Borgh:2016cco}
(see ref.~\cite{Kawaguchi:2012ii} for a review of spinor BEC), where 
the Gross-Pitaevskii equation for the latter is 
the same as the Ginzburg-Landau equation 
for a $^3P_2$ superfluid: 
a scalar BEC sandwiched between two spin-2 BECs.
The other is $^3$He superfluids with 
the A-B phase boundary 
on which one vortex in the A-phase is connected 
to two vortices in the B-phase through a boojum 
\cite{PhysRevLett.89.155301}, 
therefore an A-B-A phase configuration is relevant 
(a B-A-B phase configuration was experimentally realized 
\cite{Bradley2008}).

\subsubsection{Akaike Information Criterion (AIC)}
The Akaike Information Criterion (AIC) is defined as
\begin{equation}
\mathrm{AIC} = 2 k - 2 \ln( \hat{L}),
\end{equation}
where $\hat{L}$ is the maximum of the likelihood function $L$ (often simply called the likelihood), which  is formed from the joint probability distribution of a sample of data given a set of model parameter values; it is viewed and used as a function of the parameters given by the data sample. Generally, a lower AIC indicates a better fit.

\end{appendix}

%
%

\end{document}